\documentclass[prl,twocolumn]{revtex4} 

\usepackage{t1enc,alltt,amssymb,amsmath,natbib}
\usepackage{graphicx}

\begin{document}

\title{How wing compliance drives the efficiency of self-propelled flapping flyers}
\author{Benjamin Thiria and Ramiro Godoy-Diana}
\affiliation{{Physique et M\'ecanique des Milieux H\'et\'erog\`enes (PMMH)}\\
{UMR7636 CNRS; ESPCI ParisTech; UPMC; Universit\'e Denis Diderot}\\
{10, rue Vauquelin, F-75231 Paris Cedex 5, France}}


\begin{abstract}
Wing flexibility governs the flying performance of flapping wing flyers. Here we use a self-propelled flapping-wing model mounted on a ``merry-go-round'' to investigate the effect of wing compliance on the propulsive efficiency of the system. Our measurements show that the elastic nature of the wings can lead not only to a substantial reduction of the consumed power, but also to an increment of the propulsive force. A scaling analysis using a flexible plate model for the wings points out that, for flapping flyers in air, the time-dependent shape of the elastic bending wing is governed by the wing inertia. Based on this prediction, we define the ratio of the inertial forces deforming the wing to the elastic restoring force that limits the deformation as the \emph{elasto-inertial number} $\mathcal{N}_{ei}$. Our measurements with the self-propelled model confirm that it is the appropriate structural parameter to describe flapping flyers with flexible-wings.
\end{abstract}

\maketitle

Flapping flight is probably the way of locomotion using the most complex dynamics in the animal realm \cite{Dudley2000book,Alexander2004book}. Structural properties of animal wings, together with wing kinematics, constitute the basic elements of a tough problem: being able to stop, accelerate, execute sharp turning, hover, etc... However, flapping costs a large amount of energy due to the perpetual cycle of acceleration and deceleration involved in the process of generating useful aerodynamic forces \cite{Norberg2002}. As a means of minimizing this cost, natural systems have presumably optimized animal wings by tuning their flexibility, which enhances not only their mechanical resistance, but also the animals' flight efficiency.
The crucial nature of the elastic response of the wings in the propulsive performance of a flapping flyer has been made clear not only by observing natural systems \cite{Norberg2002}, but also investigating simplified models where wing compliance determines drastic changes in thrust production and efficiency (see \cite{Katz78,Liu97,Prempraneerach04,Miao2006,Heathcote2008}). A few ways by which wing flexibility is favorable for both flying animals and man-made devices have recently been proposed \cite{Shyy2008book,Vanella2009,Young2009} (see also extensive review by Shyy \textit{et al.} \cite{Shyy2010}).  However, although a common hand waving argument is that wing compliance can be beneficial for the flapping flyer if elastic potential energy can be stored when the wings bend and released in a favorable part of the flapping cycle, the details of the balance of fluid dynamical, structural and inertial forces and moments that governs these mechanisms remain not well understood. Experimentally, one has to note that most studies do not consider self-propelled objects but the interaction between flapping bodies held static and an oncoming uniform flow that is driven independently of the flapping motion. If one thinks of cruising flapping flight, the very fact of decoupling the flapping dynamics and the forward speed makes it difficult to extrapolate any conclusions about flight performance to the case of a free flying animal or machine. A notable exception is the experiment by Vandenberghe, Zhang and Childress \cite{Vandenberghe2004}, where a heaving wing mounted on an axis free to rotate was shown to spontaneously give rise to a cruising speed perpendicular to the direction of the heaving motion. This work has been recently extended by introducing a pitching degree of freedom to mimic wing compliance \cite{Spagnolie10}.
\begin{figure}
\centering
\includegraphics[width=0.8\linewidth,clip=true]{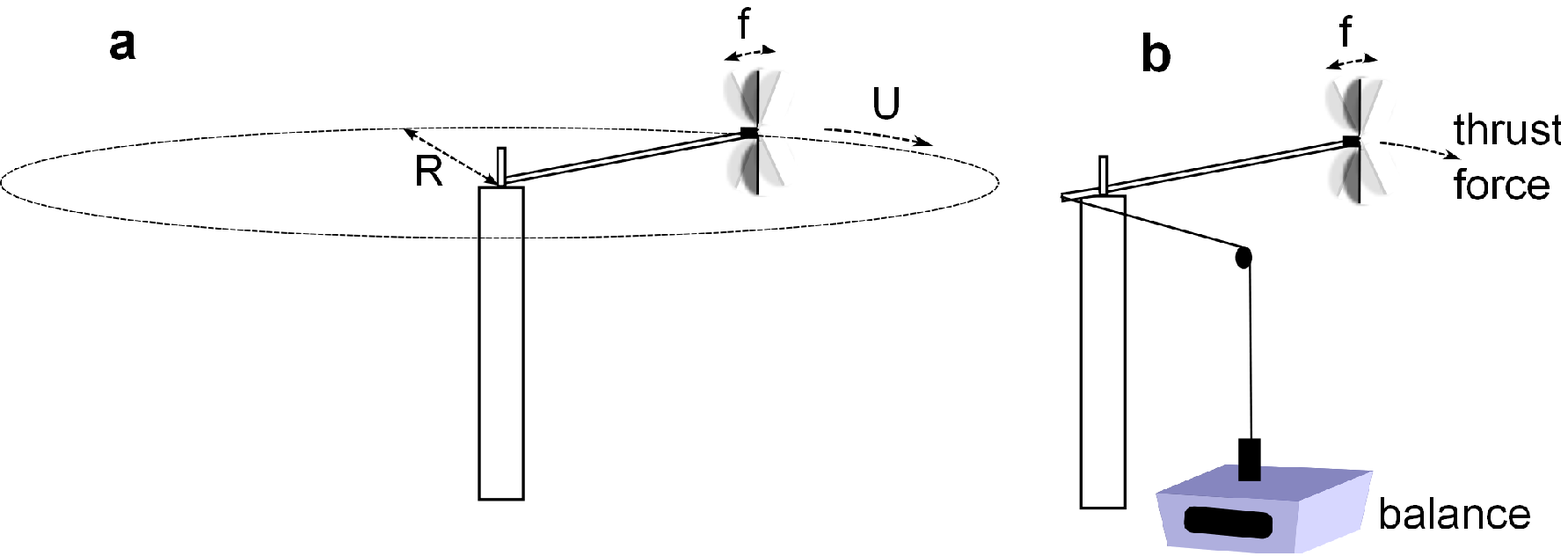}
\includegraphics[width=0.9\linewidth,clip=true]{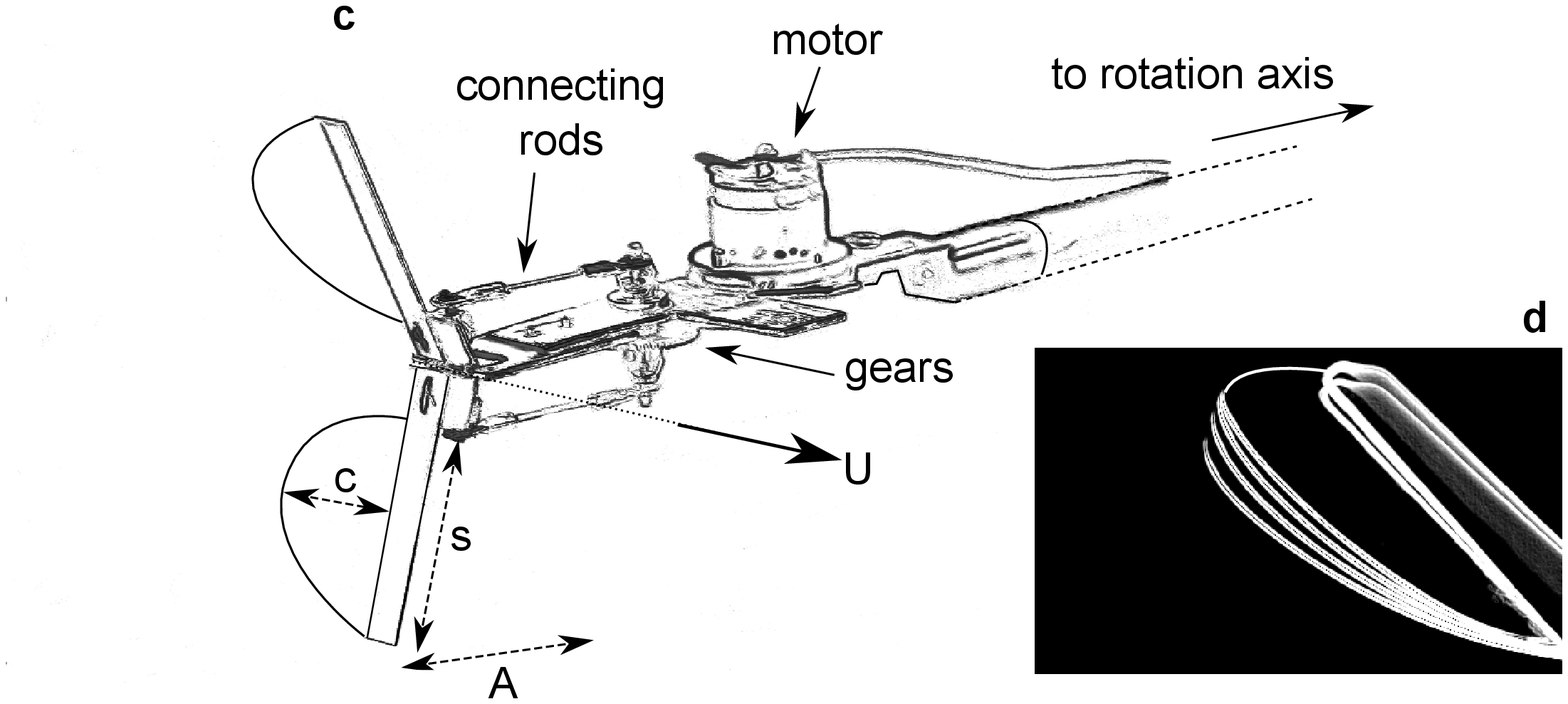}
\caption{Experimental setup. (a) Sketch of the ``merry-go-round''. The wings are mounted perpendicularly to the radial mast (length $R=0.5$m). (b) Force measurement with a tabletop balance. (c) Close up view of the flapping device. (d) Superposed pictures of the flexible flapping wing showing the chord-wise deformation. The semicircular poly-vinyl-chloride (PVC) wings had a $c=30$mm maximum chord (at mid-span). A fiberglass structure attaches each wing to the flapping mechanism and rigidifies the leading edge. The peak-to-peak flapping amplitude $A$ at the span position of the maximum chord was set to $A=30$mm. }
\label{fig_setup}
\end{figure}

Here we use an experimental self-propelled flapping-wing model to study the effect of wing flexibility on its propulsive performance. The setup has been designed to enable measurements of the cruising speed, the thrust force, as well as the consumed power, as a function of the imposed wing motion (flapping frequency) and wing design. It is shown that increasing wing flexibility leads\begin{widetext}

\begin{figure}
\includegraphics[width=0.31\textwidth]{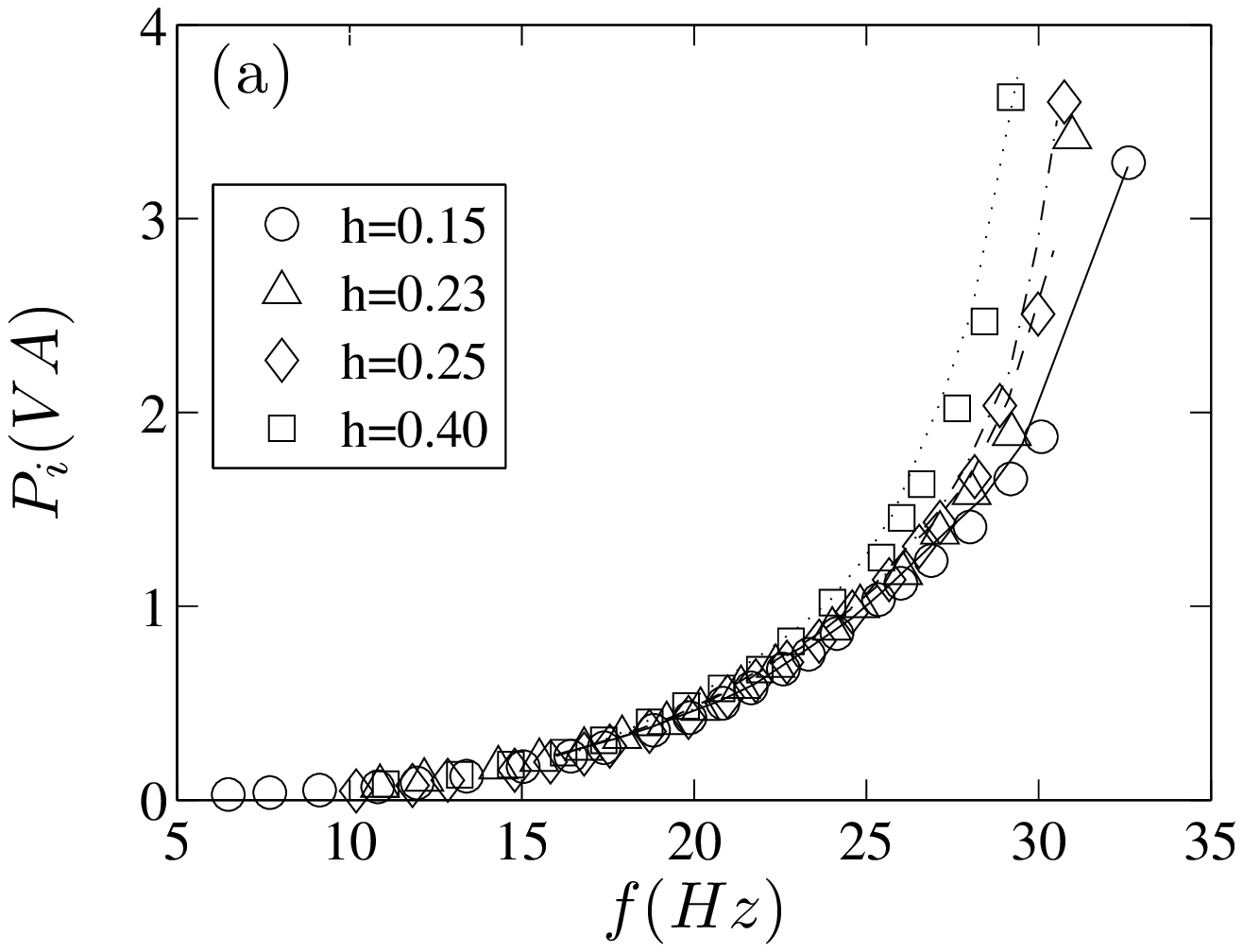}
\includegraphics[width=0.31\textwidth]{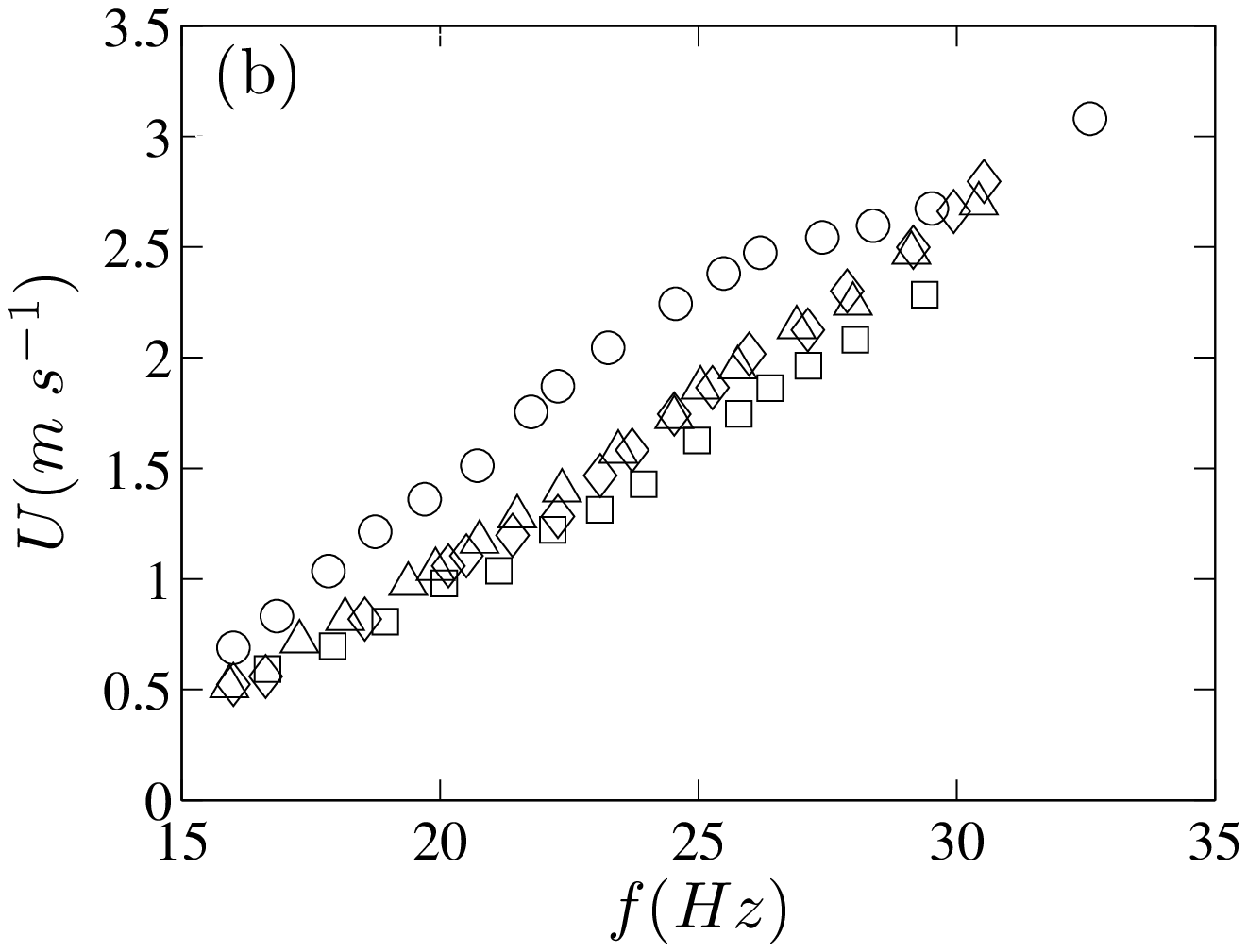}
\includegraphics[width=0.31\textwidth]{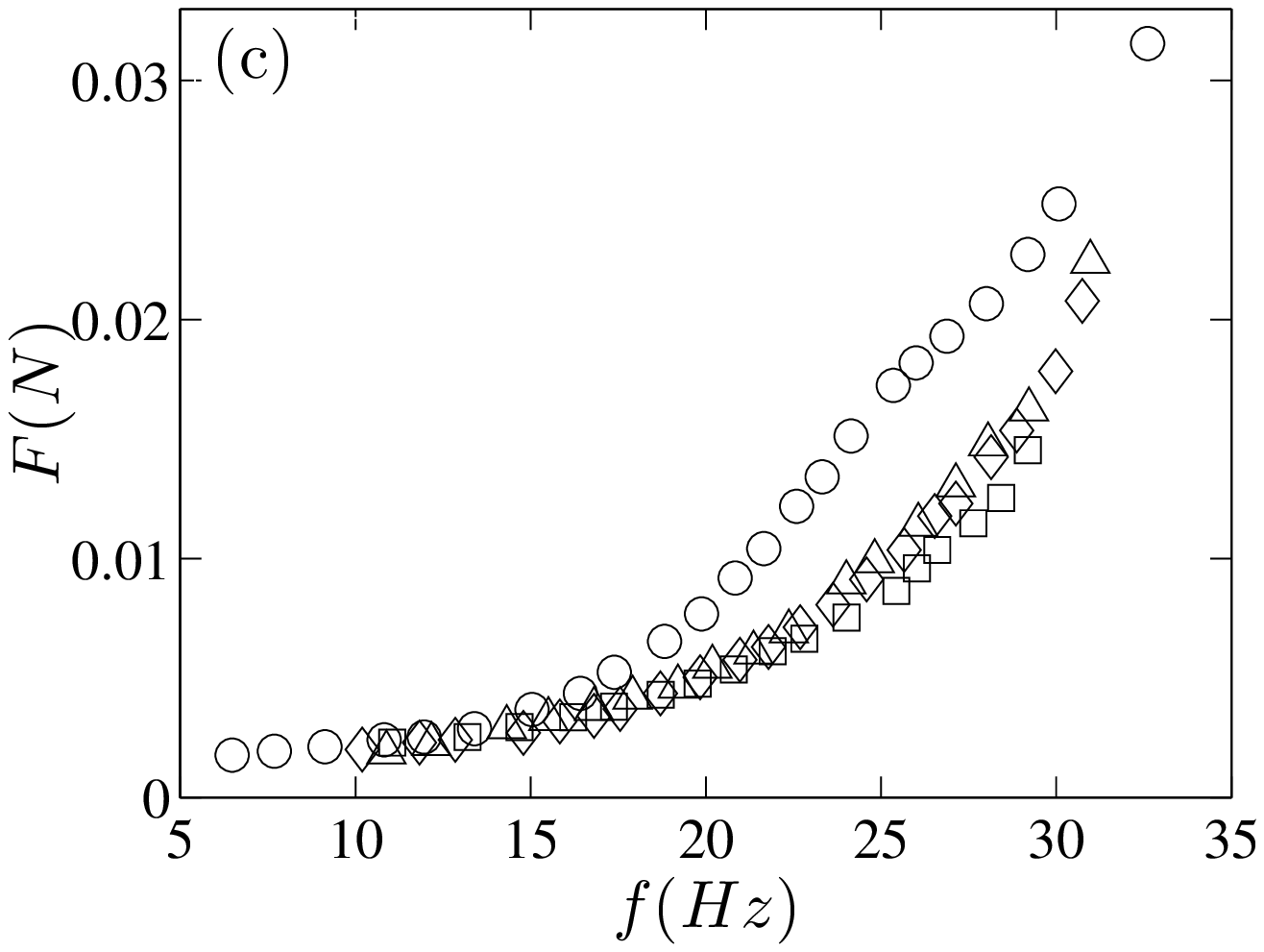}
\caption{(a) Electrical power consumed by the motor (b) cruising velocity and (c) thrust force as a function of the flapping frequency for the four tested pairs of wings. In (a) the points correspond to the force measurements with the system held at a fixed station, whereas the lines are the measurements taken with the system turning at its cruising speed.}
\label{Results}
\end{figure}

\end{widetext}
 to a substantial reduction of the consumed power as well as an increase of the thrust power, both quantities being directly dependent on the flexural properties of the wings.

The experimental setup consists of a flapping-wing device that is allowed to turn on a ``merry-go-round" type base (in the spirit of \textit{le petit man\`ege} from Marey \cite{Magnan1934}). The two-wing flapper is attached to a mast that is ball-bearing mounted to a central shaft in such a way that the thrust force produced by the wings makes the flapper turn around this shaft (see figure \ref{fig_setup}). For the chosen wing geometry (half disk of diameter $S=2L=6 $ cm), the control parameters of the experiment are the flapping frequency $(f)$ and the foil chord-wise flexibility (governed by its thickness $h$). Four pairs of wings were tested, of thicknesses 0.15, 0.23, 0.25 and 0.4 mm that correspond to masses per unit area $(\mu_s)$ of 0.20, 0.30, 0.33 and 0.53 kg m$^{-2}$ and bending rigidities ($B$) of 6, 25, 34 and 120 mN m, in the range of the flexural stiffness of real insect wings \cite{combes2003}. The natural frequencies of oscillation of the wings measured with relaxation tests are 71, 111, 125 and 166 Hz, respectively. The measured quantities are the power consumption ($P_i$, computed from voltage and current measurements in series \cite[see also][]{buchholz2008}) from which the power of the system running with no wings has been subtracted, the cruising speed of the device ($U$ obtained using the measured time per revolution $T=2\pi\Omega^{-1} $) and the thrust force $F_{T}$. The force measurement was performed by holding the flapper in a fixed station using a string attached through a pulley to a calibrated weight as shown in figure \ref{fig_setup} (top right) and monitoring the weight deficit on a tabletop balance as a function of the flapping frequency. The experiments reported here were performed with flapping frequencies ranging from 10 to 30 Hz, which drove cruising speeds between 0.2 and 1.5 m s$^{-1}$. Thus, the chord-based Reynolds number $Re=Uc/\nu$ (where $c$ is the maximum chord and $\nu$ is the kinematic viscosity) ranged between 1000 and 3000 whereas the amplitude-based Strouhal number usually used to characterize flapping-based propulsive systems $St_A=fA/U$ ranged between 0.3 and 0.9 (using the flapping amplitude at mid-span to define $A$). These values of $Re$ and $St_A$ are in the ranges that correspond to flapping flight in nature.

\begin{figure}
\includegraphics[width=0.48\linewidth]{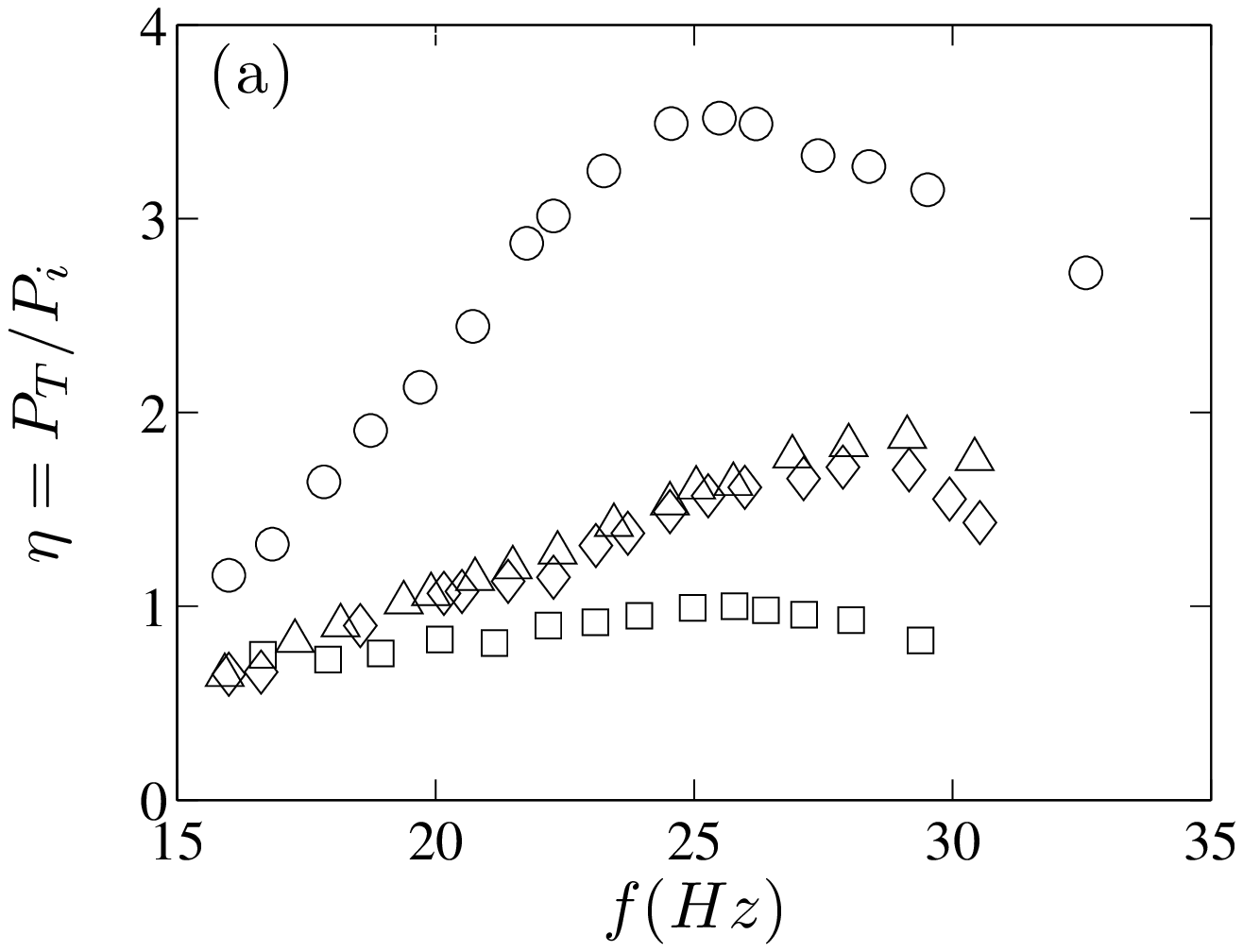}
\includegraphics[width=0.48\linewidth]{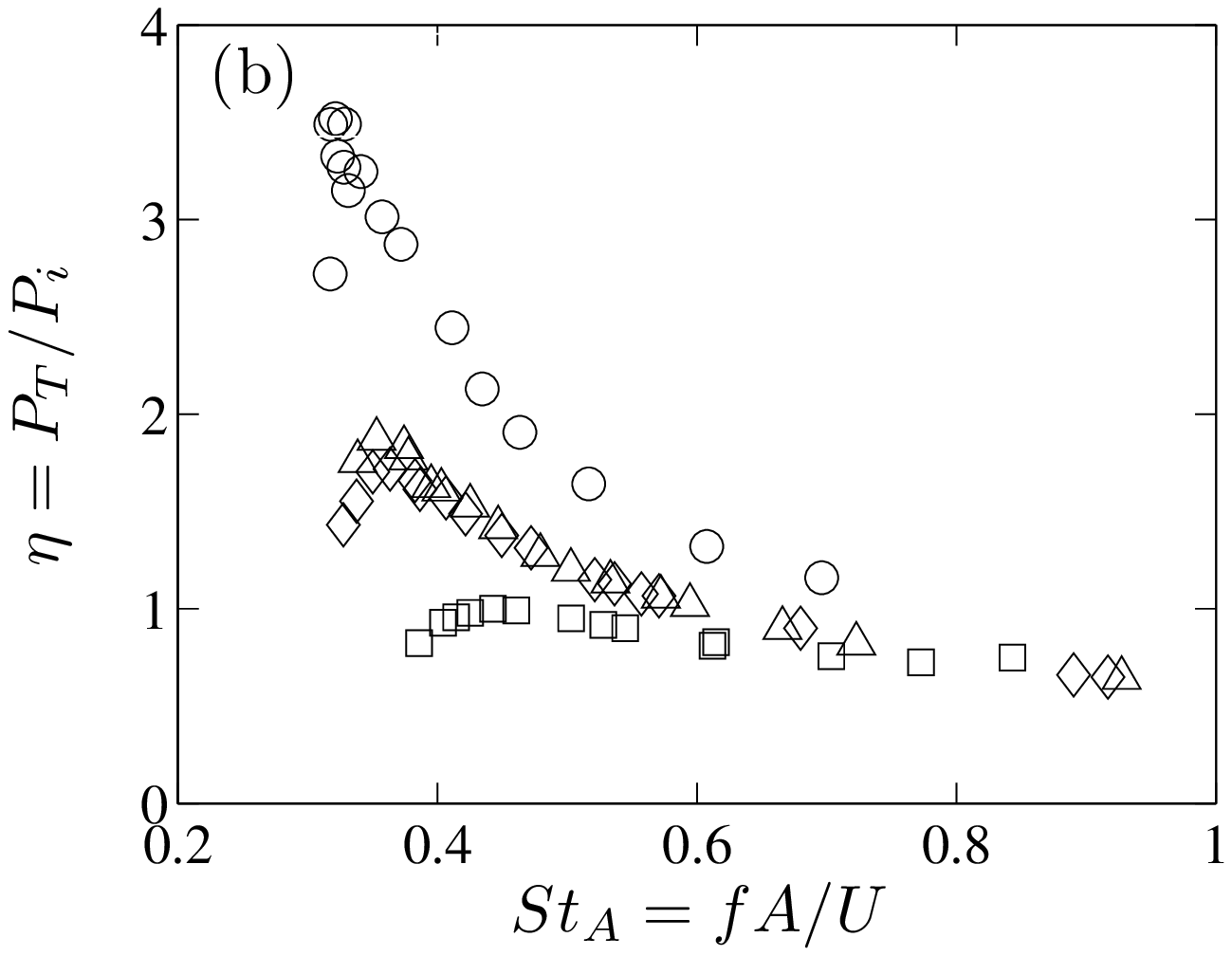}%
\caption{Normalized efficiency factor $\eta$ as a function of (a) frequency and (b) Strouhal number $St_A=fA/U$. The efficiency factor $\tilde{\eta}=P_T/P_i$ has been normalized as $\eta=\tilde{\eta}/\max{\tilde{\eta}_{\mathrm{rigid}}}$.}
\label{fig_eta}
\end{figure}

The measurements are summarized in Fig. \ref{Results}.  First, the electrical power consumed by the motor $P_{i}$ as a function of the flapping frequency $f$ is shown in Fig. \ref{Results} (a). As can be observed, keeping the system at a certain frequency requires less power as the wings become more flexible. It is also worth noting that the consumed power $P_{i}$ as a function of the forcing frequency does not change whether the system is running (lines) or not (points), i.e. whether we are measuring the thrust force or the cruising speed. This suggests that the modification of the aerodynamic forces due the the cruising motion does not play a crucial role in the energy needed to perform the flapping. We note that the most flexible wing tested can save up to 60 $\%$ in consumed power with respect to the most rigid one in the $25-30 Hz$ frequency range. Figures \ref{Results} (b) and (c) respectively show the cruising forward flight velocity $U=R\Omega $ and the thrust force $F_{T}$ as a function of the input power. Again, for the wing rigidities tested here performance increases with increasing flexibility. In a general way, the present results show that the passive mechanisms associated with wing compliance can increase flight efficiency by increasing the thrust force and the cruising velocity while spending less energy. This efficiency enhancement is summarized in Fig. \ref{fig_eta}, where the ratio of thrust power $P_{T}=F_T \times U$ to input power $P_{i}$ is displayed. All measurements are thus combined in this \emph{efficiency factor} $\tilde{\eta}=P_T/P_i$, which has been normalized in the plots with respect to its maximum value for the most rigid wing as $\eta=\tilde{\eta}/\max{\tilde{\eta}_{\mathrm{rigid}}}$. Fig. \ref{fig_eta} (a) shows that each wing has an optimum flapping frequency beyond which the efficiency starts to decrease. It is worth noting that this optimum occurs significantly below the natural relaxation frequency so that any conclusion on the role of a resonance to minimize the cost of bending (as discussed for instance in the case of undulatory propulsion by \cite{Long1996}) is not applicable (see also \cite{Michelin09}).
The same efficiency factor is plotted in Fig. \ref{fig_eta} (b) with respect to the Strouhal number $St_A$, where now the position of the optimum decreases with wing flexibility. It occurs on a range between 0.3 and 0.45, which is consistent with the range of optimal Strouhal numbers observed in nature \cite{Taylor2003}. The $\eta$ vs. $St_A$ curves show different behaviors depending on the bending rigidity of the wing and prompt us to look for a new scaling including not only aerodynamical variables but also the wing structural parameters.

The first point to address when considering the effect of wing flexibility is to identify the main force that will bend the wing. In the dynamic regime, an elastic flapping wing is subjected to both the fluid dynamic pressure acting on the surface of the wing and the inertia force due to the oscillating acceleration. A measure of the importance of these two bending forces can be given using a simplified model for the flapping wing as a plate of length $L$, mass surface density $\mu_s$ and bending rigidity $B$ (for a plate of thickness $h$ and Young's modulus $E$, $B\sim Eh^3$) whose leading edge is heaving sinusoidally with frequency $\omega$ and amplitude $A$. The moment of the mean fluid pressure force scales then as $M_f\sim\rho_f u_f^2 L^3=\rho_f \omega^2 A^2 L^3$, where $\rho_f$ is the fluid density and $u_f=A\omega$ is the maximum flapping velocity, whereas the moment of the inertia force scales as $M_i\sim \mu_s L^3 A\omega^2$. The ratio of these two moments $\frac{M_i}{Mf}$ is actually a mass ratio $\frac{\mu_s}{\rho_f A}$, which is greater than 10 for all the wings tested in the present case. The main bending factor in this case is thus the inertia force, which will be counterbalanced by the elastic restoring force produced by the bent wing. This is consistent with the analysis by \cite{Daniel2002} who concluded for most wings moving in air that the feedback between fluid pressure stresses and the instantaneous shape of the wing is negligible with respect to the inertial-elastic mechanisms.  The results in Fig. \ref{Results} (a) also support this assumption, since the change in the aerodynamic forces in the cruising regime with respect to the fixed station operation of the system is undetectable in the consumed power vs frequency curves. We therefore proceeded to compare the moment of the inertial force $M_i$ to that of the elastic restoring force that scales as $M_e\sim B$. The ratio $M_i/M_e$, which we define as the \emph{elasto-inertial} number

\begin{figure}
\centering
\includegraphics[width=0.8\linewidth]{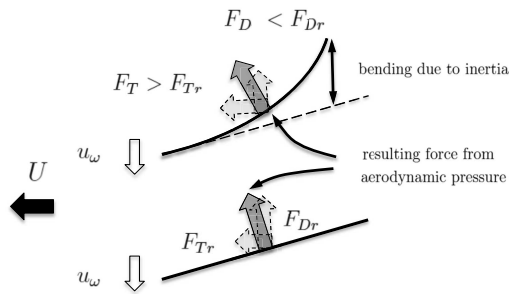}%
\caption{Schematic diagram of the redistribution of aerodynamic forces by a bending plate model (top) with respect to a rigid plate (bottom). $F_{Tr}$ is the thrust force and $F_{Dr}$ the drag forces for a rigid wing, respectively.}
\label{SketchWing}
\end{figure}

\begin{figure}
\includegraphics[width=0.5\linewidth]{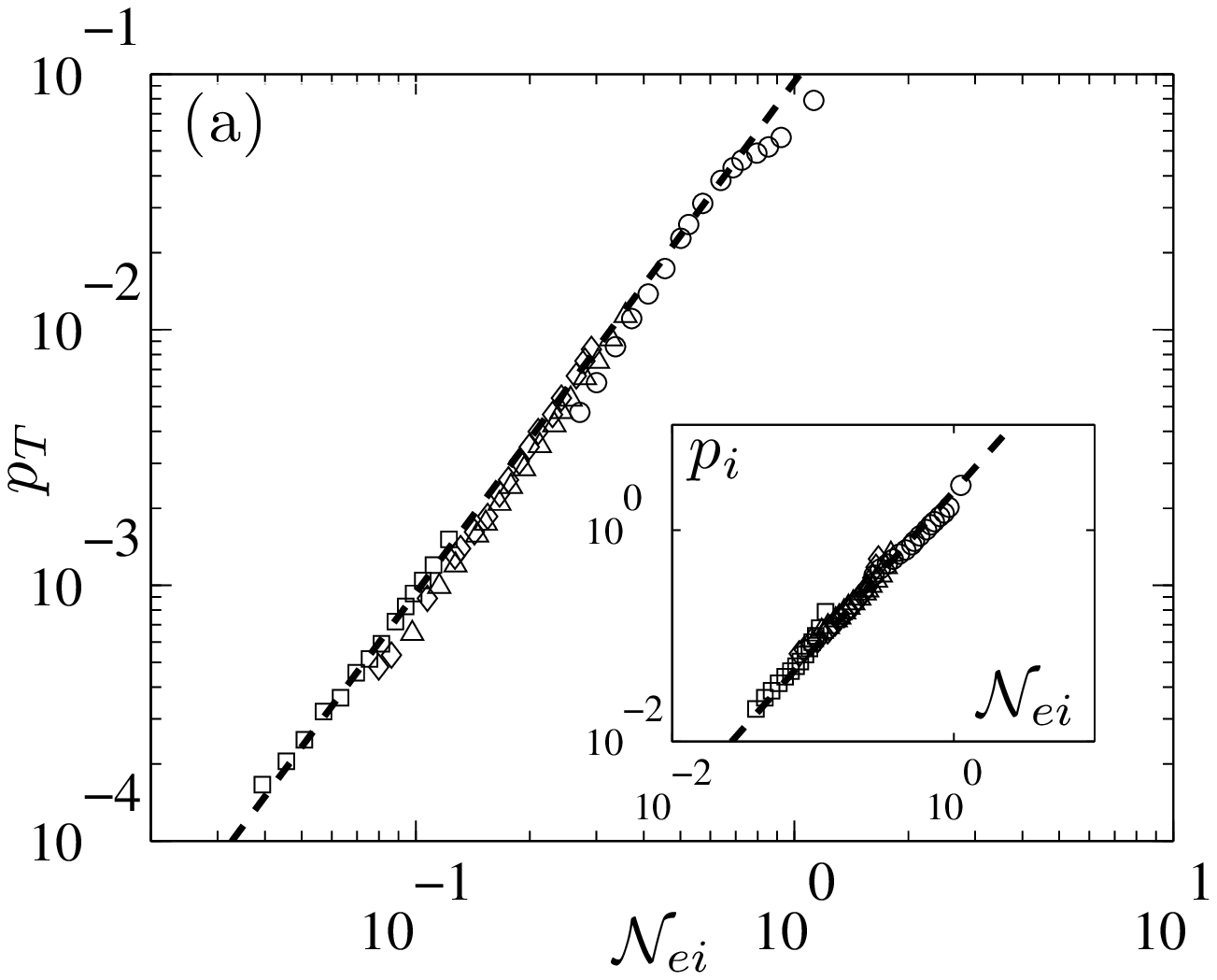}
\includegraphics[width=0.48\linewidth]{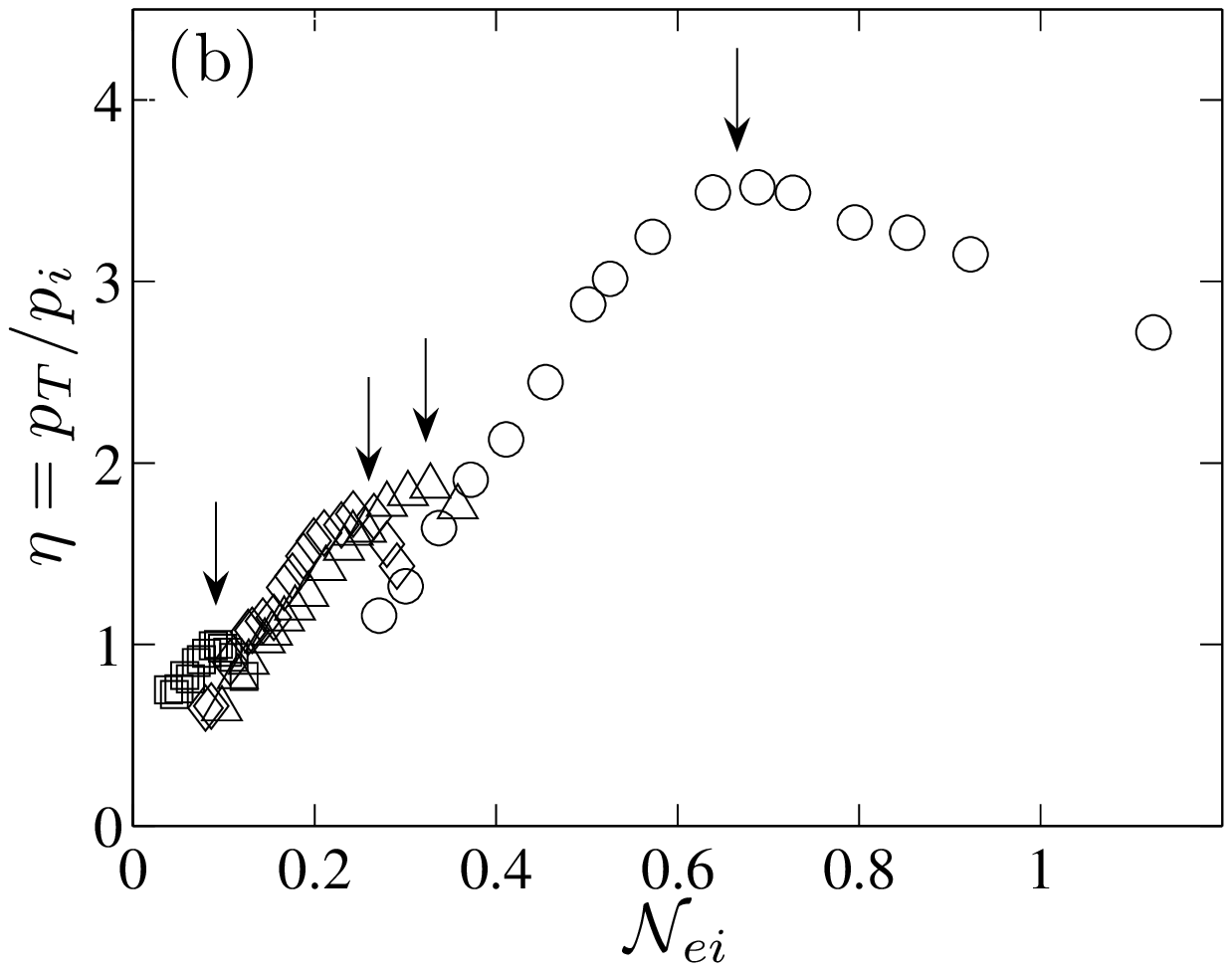}%
\caption{(a) Non-dimensional thrust power $p_{T}$ (and input power $p_{i}$ in the inset) and (b) efficiency factor as a function of elasto-inertial parameter $\mathcal{N}_{ei}$. The arrows in (b) indicate for each wing the value of $\mathcal{N}_{ei}$ where the trend of the efficiency curve changes. The error in the $\mathcal{N}_{ei}$ measurements is of 5\% due to the uncertainty in the measurements of $B$.}
\label{EffiNei}
\end{figure}

\begin{equation}\label{elastoinertial}
\mathcal{N}_{ei}=\frac{\mu_s A \omega^2 L^3}{B}=\left (\frac{L}{L_{b}} \right )^{3}\;,
\end{equation}

\noindent in terms of the \emph{bending length} $L_b=(B/\mu_s A\omega^2)^{1/3}$, measures thus to what extent the inertial force due to the oscillating acceleration will be balanced by the elastic resistance to bending (analog definitions of this bending length arise in problems where other forces drive the bending, see for instance \cite{Bico2004,Alben2002} for capillary and hydrodynamical forces, respectively). This definition determines for instance that for $\mathcal{N}_{ei}\ll 1$ the wing is too rigid for the inertia of the oscillating wing to have an observable effect, or in terms of the bending length, $L_b\gg L$ so that no deformation can be observed over the length scale $L$ of the wing chord.

Physically, the form of the bending wing can be seen as a ``shape factor'' that redistributes the contribution of the aerodynamic forces in both directions -of the flapping motion normal to the wings $F_D$ \footnote{$F_D$ is the drag force that opposes the flapping motion so that in the standard reference frame, where the thrust force is in the direction of the cruising velocity, it is actually a fluctuating lift force, not to be confused with the drag that limits the cruising velocity.} and of the forward displacement- as sketched in Fig \ref{SketchWing}. During the flapping motion, the wings experience strong drag as they push fluid up and down during a stroke cycle. Because of the flexibility of the wing, the experienced drag scales on a length depending on $L_{b}$ \cite{Alben2002,Schouveiler2006,Gosselin2010}, instead of $L$ as it would for the rigid case. On the other hand, the change in shape induces a contribution of the aerodynamic pressure load in the forward direction that is also dependent on the wing bending.
The two forces $F_{D}$ and $F_{T}$, respectively normal and in the direction of the cruising speed, should therefore be directly dependent on the wing shape that is determined by the non-dimensional number $\mathcal{N}_{ei}$ defined in Eq. \ref{elastoinertial}. This is clearly shown in Fig. \ref{EffiNei} (a), where the useful thrust power and the input power are plotted as a function of $\mathcal{N}_{ei}$. The thrust power is defined by the product of the cruising speed and the thrust force (shown in Fig. \ref{Results} (b) and (c), respectively), which are rendered non-dimensional using the scalings, $f_{T}=F_{T}L/B$ and $u=U/A\omega$. The non-dimensional thrust power is thus defined as $p_{T}=UF_{T}L/BA\omega$. Fig. \ref{EffiNei} (a) clearly shows that all the measured data collapse onto a single power law with $\mathcal{N}_{ei}$. The input power (shown in the inset of Fig. \ref{EffiNei} (a)) is non-dimensionalized using the same scaling $p_{i}=P_{i}L/BA\omega$, and gives a measure of the work of the drag force experienced by the flapping (i.e. $\sim F_{D} u_{\omega}$). This quantity also scales with the elasto-inertial number $\mathcal{N}_{ei}$. The efficiency (plotted in Fig. \ref{EffiNei} (b) as a function of $\mathcal{N}_{ei}$) that we have defined previously as $P_T/P_i$ is thus a ratio of works: the work of the useful thrust force $F_{T}$ over the work of the drag force experienced by the flapping wing $F_{D}$. A salient feature of the $\eta$ vs. $\mathcal{N}_{ei}$ plots in figure \ref{EffiNei} (b) is that the main trend followed by the experimental points over different sets of wings is valid only up to a certain threshold for each wing after which the measured data show a change of regime. We have noted in figure \ref{fig_eta} that a simple resonance at the relaxation frequency cannot explain the observed behavior. The possible role of subharmonic resonances in the efficiency curves involving a detailed study of the phase dynamics is the subject of ongoing work.

In summary, the effect of wing flexibility on the efficiency of flapping flyers can be thought of as a two-step process: a solid mechanics problem, where the balance between inertial and elastic forces determines the instantaneous shape of the flexible wings, followed by a fluid dynamics problem, where the boundary conditions set by the previous step govern the distribution of aerodynamic forces. This simple passive mechanism, shown here to be well described using the elasto-inertial number $\mathcal{N}_{ei}$, can bring a two-fold advantage: decreasing the energy cost while enhancing the thrust power. The self-propelled system described here gives a framework that should be useful to pursue further studies on the effect of structural and geometrical properties of the wings in the performance of flapping-based propulsion.

\textbf{Acknowledgements} We thank D. Pradal for his help in the design and construction of the experimental setup, J. Bico, B. Roman and O. Doaré for discussing the problem and the French Research Agency for support through project ANR-08-BLAN-0099.

%

\end{document}